# Synergy in the Knowledge Base of U.S. Innovation Systems at National, State, and Regional Levels: The Contributions of High-Tech Manufacturing and Knowledge-Intensive Services

Loet Leydesdorff,*[a] Caroline S. Wagner,[b] Igone Porto-Gomez,[c]
Jordan A. Comins,[d] and Fred Phillips[e]

**Abstract**

Using information theory, we measure innovation systemness as synergy among size-classes, zip-codes, and technological classes (NACE-codes) for 8.5 million American companies. The synergy at the national level is decomposed at the level of states, Core-Based Statistical Areas (CBSA), and Combined Statistical Areas (CSA). We zoom in to the state of California and in more detail to Silicon Valley. Our results do not support the assumption of a national system of innovations in the U.S.A. Innovation systems appear to operate at the level of the states; the CBSA are too small, so that systemness spills across their borders. Decomposition of the sample in terms of high-tech manufacturing (HTM), medium-high-tech manufacturing (MHTM), knowledge-intensive services (KIS), and high-tech services (HTKIS) does not change this pattern, but refines it. The East Coast—New Jersey, Boston, and New York—and California are the major players, with Texas a third one in the case of HTKIS. Chicago and industrial centers in the Midwest also contribute synergy. Within California, Los Angeles contributes synergy in the sectors of manufacturing, the San Francisco area in KIS. Knowledge-intensive services in Silicon Valley and the Bay area—a CSA composed of seven CBSA—spill over to other regions and even globally.

**Keywords**: innovation, geography, states, USA, synergy, system

(8483 words)

[a] * corresponding author; Amsterdam School of Communication Research (ASCoR), University of Amsterdam
PO Box 15793, 1001 NG Amsterdam, The Netherlands; loet@leydesdorff.net
[b] John Glenn College of Public Affairs, The Ohio State University, Columbus, Ohio, USA, 43210;
wagner.911@osu.edu
[c] University of the Basque Country (EHU-UPV), Engineering School of Bilbao, Alameda Urquijo s/n. 48013 –
Bilbao, Spain; igone_porto001@ehu.eus
[d] Social and Behavioral Sciences Department, The MITRE Corporation, McLean, VA, United States;
jcomins@gmail.com
[e] University of New Mexico, Anderson School of Management, Albuquerque NM 87131, USA, and State University
of New York, Department of Technology & Society, Stony Brook, NY 11794-4404, USA;
fred.phillips@stonybrook.edu

**Introduction**

The metaphor of "national innovation systems" (NIS) induces the question of whether innovation systems are nationally organized? (Carlsson, 2006). Innovation dynamics do not honor national borders, nor are innovation opportunities limited to cities (Florida, 2002; Jacobs, 1961; Storper *et al.*, 2015) or regions (Cooke, 2002). As a model of innovation systems, however, NIS combines the ideas that innovation is systemic (Lundvall, 1988) and that innovation systems are evolving (Nelson, 1993), organized institutionally, and therefore susceptible to government policies at national levels (Freeman, 1987). Thus, the perspectives of policy analysis, institutional analysis, and (neo)evolutionary theorizing are combined.

The delineation of innovation systems in institutional terms offers the advantage of compatibility with (e.g., national) statistics (Griliches, 1994). However, an institutional perspective on innovation leads to a theory of entrepreneurship (Casson, 1997) rather than accounting for the relational dynamics of communication and innovation, which is at the core of what one seeks to measure (Carter, 1996; Godin, 2006). Using a relational perspective on innovation, the emphasis has increasingly been on co-evolutions between regional economics, economic geography, and technological options (Audretsch & Feldman, 1996; Boschma, Balland, & Kogler, 2014; Feldman & Storper, 2016). This literature suggests a mutual shaping among the various factors of knowledge production inducing trajectories and niches (Geels & Schot, 2007).

In this study, we propose a methodology that combines a relational with a positional (e.g., geographical) perspective to test the assumption of systemness at national, state, and regional levels by using interactions among the geographical, technological, and organizational distributions of companies at different levels or sectors. Storper (1997, at pp. 26 ff.) considered the mutually reflexive relations among these three dimensions as a "Holy Trinity" in regional development. The distributions of these relations, however, can be systemic to varying degrees.

Our "Triple-Helix" methodology is based on entropy statistics and thus rooted in evolutionary systems theory. Synergy can be measured as negative entropy. Leydesdorff & Ivanova (2014) showed that negative information in a Triple-Helix configuration finds its origin in redundancies that are generated when uncertainty is selected from different perspectives (Leydesdorff & Ahrweiler, 2014). New options can be generated in interactions among selection mechanisms. The total number of options—the maximum entropy—is thus increased. The increase in the redundancy may outweigh the increase of uncertainty generated in ongoing processes of variation. Additional redundancy reduces relative uncertainty by adding options to the system. Increasing the number of options may be more important for the viability of an innovation system than the options realized hitherto (Fritsch, 2004; Petersen *et al.*, 2016). Furthermore, reduction of uncertainty can be expected to improve the climate for investments (e.g., Freeman & Soete, 1997, pp. 242 ff.).

We assume that three different dynamics—industrial, R&D, and political—are operating selectively upon one another. While two selection mechanisms can be shaped mutually along a trajectory, a complex dynamics is generated when three or more subdynamics interact. A third variable, for example, may make a correlation between the other two spurious. A triangle of relations can rotate clockwise in terms of feedforwards or counter-clockwise in terms of feedbacks (Ulanowicz, 2009). Feedforwards can make a system prosperous, while with the opposite sign, hyper-selectivity may lead to lock-ins and historical stagnation (Bruckner, Ebeling, Montaño, & Scharnhorst, 1996).

From this perspective, the national, regional, or sectorial levels can be considered as specific integrations among the (sub)dynamics (Carlsson, 2006). Both integration in local instantiations and differentiation among the next-order (global) selection environments operate continuously in systems of innovation. The local combinations instantiate historical trajectories, while the interactions among the selection environments (markets, governance, R&D) develop at a next-order regime level. Interactions among selection mechanisms generate redundancy when the selection mechanisms overlap. Since the two processes—the historical generation of variation and the evolutionary interactions among selection environments—are operating concurrently, the trade-off between uncertainty generation and reduction can be expected to vary among regions, sectors, etc. This trade-off can be measured in bits using the TH indicator (Leydesdorff, Park, & Lengyel, 2014).

The research question thus becomes: to what extent can a given configuration such as a national or regional portfolio be expected to operate not only as a *system*, but also as an *innovation system*? A measure for systemness can easily be developed, for example, on the basis of the Markov property: the current state of a system provides a better prediction of its next state than what can be derived from the history of its elements. Using publication data, for example, Leydesdorff (2000) showed that the European Union (at that time) was evolving as a set of national research systems more than at the European level. German unification, however, led to the shaping of a single publication system in Germany during the 1990s.

An *innovation system* would not only evolve as a system, but also generate new options. Redundancy generation increases the maximum entropy. Biological systems increase uncertainty following the entropy law (Brooks & Wiley, 1986). Technological innovation, however, extends the number of options. For example, the capacity of transport across the Alps could be considered as constrained by the capacity of roads and railways such as at the Brenner Pass. As one invents new channels, however, other options became available, such as for example air transport across the Alps or tunnels underneath, which are not constrained by the geological or weather conditions on the ground.

Both redundancy and information are generated in TH-type innovation systems. The feedback and feedforward loops precondition each other: the phenotypical variation can be *organized* historically (for example, by governments and in enterprises). The selection mechanisms have the status of hypotheses; they can be considered as self-organizing "genotypes" (Hodgson & Knudsen, 2011; Langton, 1989). Unlike biological code (DNA), selection is not hard-wired but operates as a code in the communication. The selection criteria can be expected to adapt evolutionarily to the opportunities provided in the historical layer. Using the TH-indicator for the measurement of the trade-off, *positive* mutual information among the three helices indicates that the generation of (Shannon-type) information prevails; when this measure is *negative*, the non-linear generation of redundancy (in loops) prevails, and uncertainty is reduced (cf. Krippendorff, 2009).

In this study, we apply this methodology to studying the knowledge base of the American economy. We have applied the approach in a number of (mainly European) country studies.[6] However, the application to data about the U.S.A. is expected to provide new insights concerning both the effectiveness of the measurement model and the knowledge base of the U.S. economy. Our methodology enables us to test whether or specify the extent to which synergy among distributions is generated and systems can be considered as innovation systems. We focus on geographical scales, but will distinguish also in terms of sectors such as high- and medium-tech manufacturing and knowledge-intensive services (Carlsson, 2013). We thus endogenize the technological dimension into the model (Nelson & Winter, 1977).

**The American innovation system**

In a review of the U.S. innovation system and innovation policy, Shapira & Youtie (2010) argue that the U.S. system is marked by diversity and multiple layers and levels to the extent that one may question whether a national system of innovations is even a useful concept. The authors emphasize the role of the States, which they formulate (at pp. 4-5) as follows:

> State governments tend to be much more active in the innovation area than the federal government
> has been, primarily because there has traditionally been reluctance at the federal level to intervene in
> industrial policy, while state governments are closer to the needs of the particular industries that

---

[6] Italy (Cucco & Leydesdorff, 2013; in preparation); Hungary (Lengyel & Leydesdorff, 2011); the Netherlands (Leydesdorff, Dolfsma, & Van der Panne, 2006) ; Germany (Leydesdorff & Fritsch, 2006); Russia (Leydesdorff, Perevodchikov, & Uvarov, 2015); Spain (Leydesdorff & Porto-Gomez, 2017); Sweden (Leydesdorff & Strand, 2013); China ( Leydesdorff & Zhou, 2014); Norway (Strand & Leydesdorff, 2013)

> make up their regional economies. Many recent federal programs have had historic roots in long standing state and local innovation initiatives.

Innovation is concentrated in a few states: in 2009, about 67 percent of all venture capital deals and 74 percent of venture capital dollars flowed to the top five states. By 2014, those states' share of venture dollars grew to 80 percent, according to NVCA/Pricewaterhouse Coopers. R&D funds also go overwhelmingly to five states. California-based companies received about 56 percent of all U.S. venture capital dollars in 2014.[7]

We test the hypothesis of innovation-systemness at the three geographical levels of States, CBSAs, and CSAs in terms of high- and medium-tech manufacturing and knowledge-intensive services. The order of presentation is top-down; we zoom in on California and conduct a more detailed evaluation and comparison of the CBSAs of San Francisco and Los Angeles (LA) and the CSA of the Silicon Valley area as examples (Storper *et al*., 2015).

According to Audretsch & Feldman (1996) and many other authors, Silicon Valley has been the region with the largest number of innovations, followed by New York, New Jersey, and Massachusetts. However, LA is more important in terms of high- and medium-tech manufacturing than the Bay area (Feldman & Florida, 1994), while San Francisco dominates in terms of knowledge-intensive services (Whittington *et al*., 2009). Silicon Valley provides a mixture of high-tech manufacturing and knowledge-intensive services (Bresnahan & Gambardella, 2004), but the economic activity of this region is less rooted geographically than in the other two areas (Saxenian, 1996). The more detailed analysis of California and Silicon Valley will enable us to discuss some of the limitations of the methodology.

---

[7] https://ssti.org/blog/useful-stats-share-us-venture-capital-investment-state-2009-2014

We use companies as the units of analysis and specify three codes as most relevant for innovation systems: (1) ZIP codes indicating company addresses in the geographical dimension, (2) NACE codes developed by the OECD as indicators of the technological capabilities of companies, and (3) size-classes as proxies for organizational formats such as small- and medium-sized companies versus large corporations. The data are disaggregated at the level of 51 states,[8] approximately one thousand Core-Based Statistical Areas (CBSA), and 171 so-called Combined Statistical Areas (CSA). CBSAs are defined by the U.S. Office of Management and Budget (OMB) as geographical zones of one or more counties (or equivalents) anchored by an urban center of at least 10,000 people and including adjacent counties that are socioeconomically tied to the urban center via commuting. CBSAs can be metropolitan or micropolitan (e.g., rural; Brown *et al*., 2004; Hall, 2009). CSAs can be defined (by the OMB) when multiple metropolitan or micropolitan areas have an employment interchange of at least 15%;[9] CSAs often represent regions with overlapping labor and media markets.

---

[8] The District of Columbia is included as a state.
[9] OMB Bulletin No. 17-01: Revised Delineations of Metropolitan Statistical Areas, Micropolitan Statistical Areas, and Combined Statistical Areas, and Guidance on Uses of the Delineations of These Areas, at
https://www.whitehouse.gov/sites/whitehouse.gov/files/omb/bulletins/2017/b-17-01.pdf

**Methods and data**

*Data*

Data were retrieved from the ORBIS database of Bureau van Dijk on May 4-6, 2017,[10] using the search string "United States of America" for all active companies with data covering a known value and a last available year, including estimates for the number of employees where necessary. Companies with no recent financial data were excluded, as were public authorities, states, and governments. We follow the definition and delineation of companies as provided in ORBIS. This constraint on the data is a major limitation. ZIP codes, for example, vary over geographical regions; however, in reference to the other two dimensions, the distribution of ZIP codes indicates local constraints (such as infrastructure) operating as a (non-market) selection environment.

In addition to the assignment of NACE and ZIP-codes, companies are scaled in terms of the number of their employees as a third dimension. SMEs are commonly defined in these terms. Financial turn-over is available in the ORBIS data as an alternative indicator of economic structure. However, we chose the number of employees as one can expect this number to exhibit less volatility than turn-over, which may vary with stock value and economic conjecture more readily than numbers of employees. Numbers of employees are sensitive to other activities, such as outsourcing.

---

[10] When we entered the ORBIS database again on September 27, 2018 (after the review process), the coverage had grown from approximately180,000 to 230,000 companies, of which 53,624,319 with an address in the USA.

**Table 1**: Distribution of records over years in the download

| Year | Frequency | Percent | Cumulative Percent |
|------|-----------|---------|--------------------|
| *n.a.* | 9 | .0 | .0 |
| 2013 | 128,132 | 1.5 | 1.5 |
| 2014 | 596,290 | 7.0 | 8.5 |
| 2015 | 1,038,645 | 12.2 | 20.8 |
| 2016 | 6,730,064 | 79.2 | 100.0 |
| 2017 | 99 | .0 | 100.0 |
| Total | 8,493,239 | 100.0 | |

The retrieval yields a total of 8,493,322 companies, of which 8,492,239 records were accessible for download. Table 1 shows that 79.2% of the records are from 2016. Only nine records have no valid time stamp; city names were missing in 1,253 records; state names in 820 records; ZIP codes in 8,330 records; and NACE codes were missing in 364,310 records. Records without NACE codes or ZIP codes were deleted. The resulting file—our sample—contains 8,121,301 records with valid NACE and ZIP codes (Table 2).

Geographical data

In addition to various lists made available online by the U.S. Census Bureau and the Office of Management and Budget (OMB), we used two concordance tables for ensuring that geographical records were as complete as possible: (*i*) the Missouri Census Data (available at http://mcdc2.missouri.edu/websas/geocorr2k.html) and (*ii*) ZipList5 CBSA™ (June 2017; available at https://www.zipinfo.com/products/z5cbsa/z5cbsa.htm) with 5-digit ZIP codes, Core Based Statistical Area (CBSA) codes (including Metropolitan Statistical Areas, Micropolitan Statistical Areas, and Metropolitan Divisions), city and state names, etc. The definitions of this database follow the revised MSA definitions issued by the Federal Government in July 2015. A field covering CSAs was added to this data when applicable.

Table 2: Numbers of records included in the analyses.

|  | Missing Values | Sample Size | Geographical Scale |
|---|---|---|---|
| **Sample downloaded** |  | 8,493,239 |  |
| **Zip or Nace Codes incomplete** | 371,938 | 8,121,301 | States |
| **CBSA not applicable** | 1,170,620 | 6,950,681 | CBSA |
| **CSA not applicable** | 2,681,528 | 5,439,773 | CSA |

Using the various concordance lists, all records were exhaustively matched for address information. Appendix 1 provides the distribution of these companies over U.S. states. We use the first three digits of the ZIP codes corresponding to the level of counties. The fourth and fifth digits provide more detailed postal information; the data contains 923 valid ZIP codes.

Of the 8,121,301 records, 1,171,620 could not be assigned to a CBSA and 2,681,528 not to a CSA.[11] The official number of CBSA classes is 945 (since July 2015), of which 389 are metropolitan areas and 556 micropolitan. Our data includes 997 CBSA names and another six CBSA numbers without an identification. Some CBSA names used in this data are outdated. On average, a CBSA contains 7,611 companies, but the standard deviation is considerable: 27,936. Similarly, the distribution of companies across states is heterogeneous: the average is 150,394; st.dev. = 188,069.[12] Note that CBSAs, CSAs, and States are not part of the model as (horizontally interacting) variables, but are used only as (vertically different) levels of aggregation of the units of analysis. Using the concordance file,[13] 524 CBSA could be attributed to 169 CSA in 2015,[14] among which 266 metropolitan and 258 micropolitan CSA. These 524 CSA contain 6,298,681 records. The distribution is again skewed: on average, 165 CSA contain 30,721 records of companies with a standard deviation of 58,985.5.

---

[11] We combined the CBSAs "Los Angeles-Long Beach-Anaheim, CA" and "Los Angeles-Long Beach-Santa Ana, CA" into a single CBSA with the former name, which is part of the CSA "Los Angeles-Long Beach, CA."
[12] The state attributions include Guam (GU; n = 371), Puerto Rico (PR; n = 3,911), and the Virgin Islands (VI; n = 258).
[13] Available at https://en.wikipedia.org/wiki/Combined_statistical_area.

Company classification

The classification of companies in terms of the "Nomenclature générale des Activités économiques dans les Communautés Européennes" (NACE, Rev. 2) was used for indicating the technological dimension. The NACE code is derived from the International Standard Industrial Classification (ISIC) that is used in the US. We use the NACE codes, however, in order to make the results directly comparable with previous studies.[15] The disaggregation in terms of medium- and high-tech manufacturing, and knowledge-intensive services, is provided in Table 3. The data contains 254 NACE codes (Rev. 2) at the three-digit level.[16]

---

[14] Puerto Rico was not included in this analysis.
[15] A different code is the North American Industry Classification System (NAICS) which was developed in the mid-1990s to provide common industry definitions for Canada, Mexico, and the United States. NAICS is developed on the basis of a production-oriented conceptual framework and classifies units, not activities. As a result, the structures of ISIC and NAICS are substantially different (Eurostat, 2009, p. 42).
[16] A complete index of NACE codes can be found, for example, at http://www.cso.ie/px/u/NACECoder/Index.asp .

**Table 3**: NACE classifications (Rev. 2) of high- and medium-tech manufacturing, and knowledge-intensive services.

| *High-tech Manufacturing* | *Knowledge-intensive Sectors (KIS)* |
|---|---|
| 21 Manufacture of basic pharmaceutical products and pharmaceutical preparations<br>26 Manufacture of computer, electronic and optical products<br>30.3 Manufacture of air and spacecraft and related machinery<br><br>*Medium-high-tech Manufacturing*<br><br>20 Manufacture of chemicals and chemical products<br>25.4 Manufacture of weapons and ammunition<br>27 Manufacture of electrical equipment,<br>28 Manufacture of machinery and equipment n.e.c.,<br>29 Manufacture of motor vehicles, trailers and semi-trailers,<br>30 Manufacture of other transport equipment<br>• excluding 30.1 Building of ships and boats, and<br>• excluding 30.3 Manufacture of air and spacecraft and related machinery<br>32.5 Manufacture of medical and dental instruments and supplies | 50 Water transport,<br>51 Air transport<br>58 Publishing activities,<br>59 Motion picture, video and television programme production, sound recording and music publishing activities,<br>60 Programming and broadcasting activities,<br>61 Telecommunications,<br>62 Computer programming, consultancy and related activities,<br>63 Information service activities<br>64 to 66 Financial and insurance activities<br>69 Legal and accounting activities,<br>70 Activities of head offices; management consultancy activities,<br>71 Architectural and engineering activities; technical testing and analysis,<br>72 Scientific research and development,<br>73 Advertising and market research,<br>74 Other professional, scientific and technical activities,<br>75 Veterinary activities<br>78 Employment activities<br>80 Security and investigation activities<br>84 Public administration and defence, compulsory social security<br>85 Education<br>86 to 88 Human health and social work activities,<br>90 to 93 Arts, entertainment and recreation<br><br>Of these sectors, 59 to 63, and 72 are considered *high-tech services*. |

Sources: Eurostat/OECD (2009, 2011) ; Eurostat/OECD (2011); cf. Laafia (2002, p. 7) and Leydesdorff *et al*. (2006, p. 186).

Small, medium-sized, and large enterprises

As noted, we use the number of employees as a proxy for size of the company (Table 4). Small and medium-sized companies (etc.) are commonly defined in terms of numbers of employees. However, the definitions of small and medium-sized businesses versus large enterprises vary among world regions. Most classifications use six or so categories for summary statistics. We use the eleven classes provided in Table 3 because this finer-grained scheme produces richer results (Blau & Schoenherr, 1971; Pugh, Hickson, & Hinings, 1969a and b; Leydesdorff,

Dolfsma, & Van der Panne, 2006; Leydesdorff & Porto-Gomez, 2017; Rocha, 1999). Note that micro-enterprises (with fewer than five employees) constitute 66.3% of the companies under study.

**Table 4**. Size distribution of the companies in the sample according to the number of employees. (Source: ORBIS data.)

|  | Frequency | Percent | Valid Percent | Cumulative Percent |
|---|---|---|---|---|
| 0 or 1 | 12303 | .2 | .2 | .2 |
| 2–4 | 5374768 | 66.2 | 66.2 | 66.3 |
| 5–9 | 4819 | .1 | .1 | 66.4 |
| 10–19 | 1330453 | 16.4 | 16.4 | 82.8 |
| 20–49 | 687156 | 8.5 | 8.5 | 91.2 |
| 50–99 | 446597 | 5.5 | 5.5 | 96.7 |
| 100–199 | 153212 | 1.9 | 1.9 | 98.6 |
| 200–499 | 84946 | 1.0 | 1.0 | 99.7 |
| 500–749 | 26467 | .3 | .3 | 100.0 |
| 750–999 | 420 | .0 | .0 | 100.0 |
| > 1,000 | 160 | .0 | .0 | 100.0 |
| Total | 8121301 | 100.0 | 100.0 |  |

*Statistics*

Using Shannon's (1948) information theory, uncertainty in the distribution of a random variable $x$ can be defined as $H_x = -\sum_x p_x \log_2 p_x$. The values of $p_x$ are the relative frequencies of $x$: $p_x = f_x / \sum_x f_x$. When base two is used for the logarithm, uncertainty is expressed in bits of information.

The uncertainty in the case of a system with two variables can be formulated analogously as

$$H_{xy} = -\sum_x \sum_y p_{xy} \log_2 p_{xy} \qquad (1)$$

In this case of two variables with interaction, the uncertainty of the system is reduced by mutual information $T_{xy}$ as follows:

$$T_{xy} = (H_x + H_y) - H_{xy} \qquad (2)$$

One can derive (e.g., McGill, 1954; Yeung, 2008, pp. 59f.) that in the case of three dimensions, mutual information corresponds to:

$$T_{xyz} = H_x + H_y + H_z - H_{xy} - H_{xz} - H_{yz} + H_{xyz} \qquad (3)$$

Eq. 3 can yield negative values and is therefore not a Shannon-type information (Krippendorff, 2009). Shannon-type information measures variation, but this negative entropy is generated by next-order loops in the communication; for example, when different codes interact as selection environments.

Note that uncertainty is implicated by the variation in historical *relations*. From an evolutionary perspective, the historical networks of relations function as retention mechanisms. Our measure, in other words, does not measure action (e.g., academic entrepreneurship) or output, but the investment climate as a structural consequence of *correlations* among distributions of relations; the correlations can be spurious. However, the distinction between the structural dynamics and the historical dynamics of relations is analytical. The two layers reflect each other in the events. Eq. 3 models this trade-off between variation and selection as positive and negative contributions to the prevailing uncertainty. The question of systemness can thus be made empirical and amenable to measurement.

In the case of groups (or subsamples), one can decompose the information as follows: $H = H_0 + \sum_G \frac{n_G}{N} H_G$ (Theil (1972, pp. 20f.). The right-hand term ($\sum_G \frac{n_G}{N} H_G$) provides the average uncertainty in the groups and $H_0$ the additional uncertainty in-between groups. Since $T$ values are decomposable in terms of $H$ values (Eq. 3), one can analogously derive (Leydesdorff & Strand, 2013, at p. 1895):

$$T = T_0 + \sum_G \frac{n_G}{N} T_G \qquad (4)$$

In this formula, $T_G$ provides a measure of uncertainty at the geographical scale $G$; $n_G$ is the number of companies at this scale, and $N$ is the total number of companies under study. One can also decompose across regions, in terms of company sizes, or in terms of combinations of dimensions.

Because the scales are sample-dependent, one may wish to normalize for comparisons across samples, for example as percentages. After normalization, the geographical contributions of regions or states can be compared in bits (or other measures) of information. In this design, the between-group term $T_0$ provides us with a measure of what the next-order system (e.g., the nation) adds in terms of synergy to the sum of the regional systems or states. The three dimensions are the (g)eographical, (t)echnological, and (o)rganizational; synergy will be denoted as $T_{GTO}$ and measured in millibits with a minus sign.

**Results**

*Decomposition in terms of U.S. states*

First we decompose the U.S. in terms of its 50+ states. Figure 1 shows the percentages of synergy contributions of states.

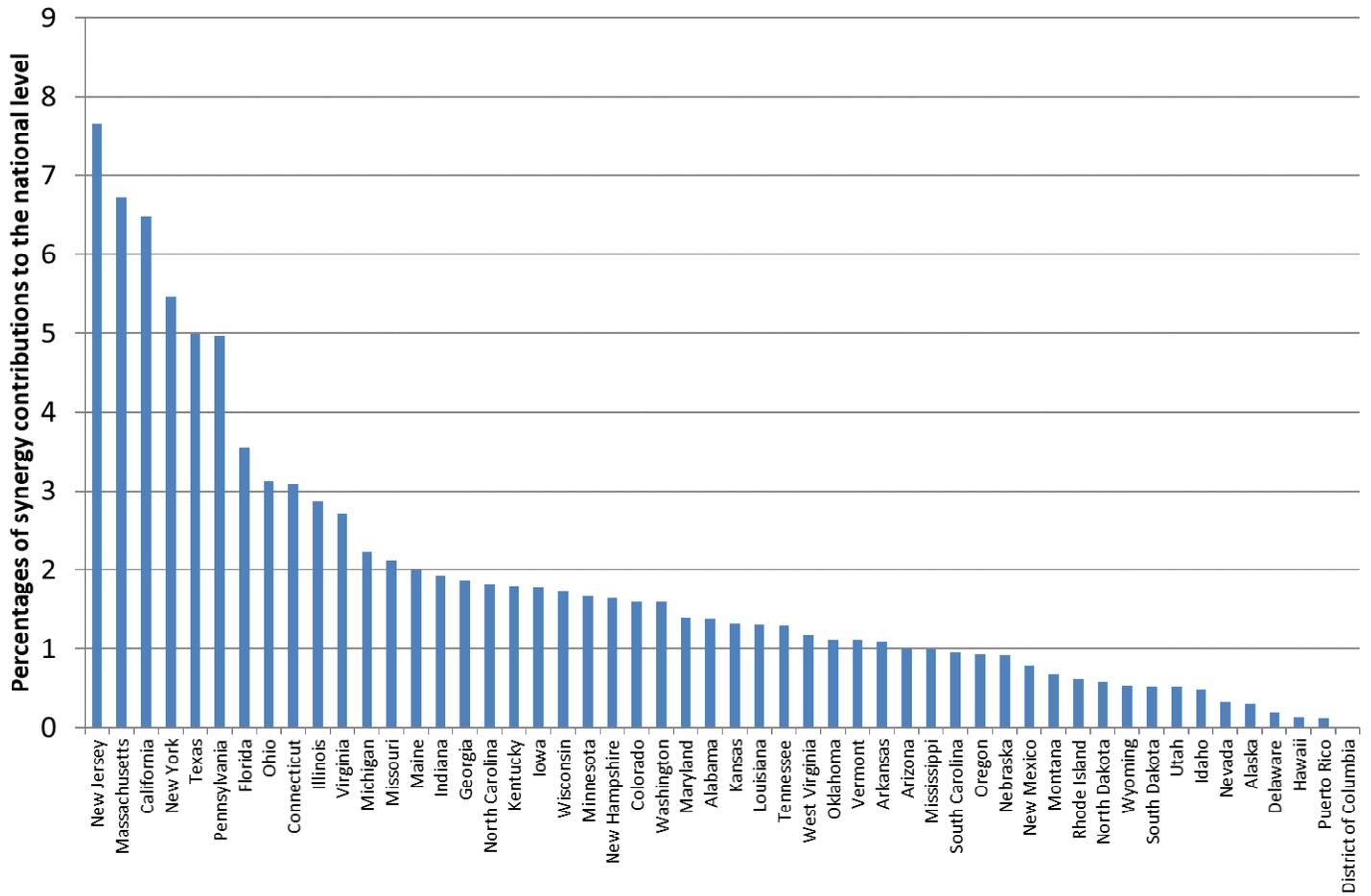

**Figure 1**: Percentages of synergy contributions of 50+ U.S. states; *n* = 8,121,301 companies.

Six states stand out as generating 36.3% of the national synergy: New Jersey, Massachusetts, New York, and Pennsylvania at the East Coast, California at the West Coast, and Texas in the south. Feldman & Florida (1994) already noted that New Jersey is the state with the largest number of innovations per worker in the manufacturing sector. The aggregation of all states accounts for 97.2% of the national synergy. Consequently, the additional synergy at the national (above-state) level is only 2.8%. This is much less than we found in previous studies of national innovation systems: Norway (11.7%), China (18.0%), the Netherlands (27.1%), Sweden (20.4%), and Russia (37.9%). In other words, the national level does not add much to the synergy at the level of the states. However, 18 states contribute less than 1% to the national synergy. The assumption of a national innovation system in the U.S. is therefore not supported by our results. We proceed in the next section with the sector-based decomposition. Does one find similar patterns when focusing on high-tech manufacturing or knowledge-intensive services? Or do we observe specialization among states and regions?

*Sectorial decomposition at the level of states*

As noted, $\Delta T$ values can be compared as percentages of contributions to the national synergy after normalization. Let us compare the four sectors specified in Table 2: high-tech manufacturing (HTM), medium-high-tech manufacturing (MHTM), knowledge-intensive services (KIS), and high-tech KIS (HTKIS) (Table 5). Note that HTKIS is a subset of KIS.

**Table 5**: Correlations of percentages of contributions over 51 states. Pearson and Spearman (rank-order) correlations in the lower and upper triangle, respectively; all correlations are significant at the 1% level; $N = 51$.

|         | % all | % htm | % mhtm | % htkis | % kis |
|---------|-------|-------|--------|---------|-------|
| % all   | 1     | .919  | .950   | .975    | .994  |
| % htm   | .974  | 1     | .966   | .946    | .900  |
| % mhtm  | .971  | .966  | 1      | .975    | .937  |
| % htkis | .992  | .968  | .973   | 1       | .968  |
| % kis   | .997  | .965  | .952   | .986    | 1     |

The high correlations in Table 5 lead to the conclusion that the distributions over the states for the various sectorial decompositions are not significantly different from one another or from the overall distribution of the synergy over the states. Thus, the synergy contribution is state-specific; the sectors only modulate the state-specific averages.

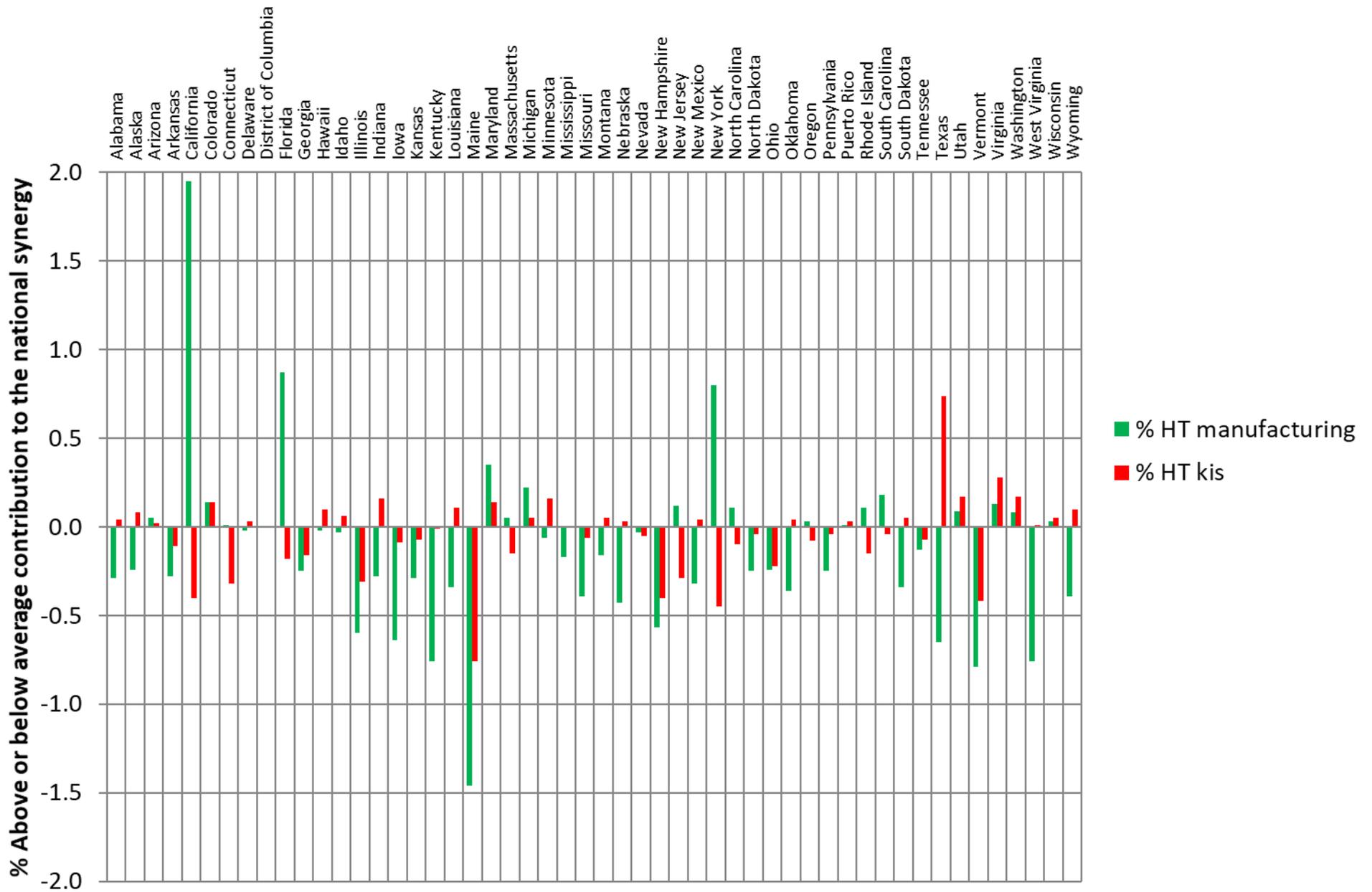

**Figure 2**: Specialization patterns contrasting HTM with HTKIS for 50+ states.

By focusing on the differences between sectorial contributions, one can visualize specializations among the states. Figure 2 shows the relative percentages of contributions to the national synergy by HTM and HTKIS for the various states when compared with the average of all records with an address in these states. Whereas only 59,621 companies (0.7%) are classified as high-tech manufacturing (HTM), companies generate 280.3 mbits of synergy. After normalization this is 2.4% of the national synergy in all sectors. Of this synergy in HTM, 90.7% occurs at the level of the states, and 9.3% is generated above the state level.

The strong position of California (green bar in Figure 2) is unambiguous in high-tech manufacturing. California also holds the strongest position in the domain of medium-high-tech manufacturing (MHTM). However, a number of older industrial states (Ohio, Indiana, etc.) also score above average on MHTM. MHTM adds to the synergy in almost all states above expectations (based on the average contribution for the state). Conversely, KIS provides less synergy than the average. KIS is less geographically rooted, since services can easily be provided across state borders.

Knowledge-Intensive Services (KIS) do not contribute synergy at the national level after aggregating the synergies at the state level. The overshoot (101.5%) indicates that a component of KIS is independent of geographical location. With 34.3% of the companies, KIS generates 29.7% of the synergy at the national level (n = 2,789,295). Within KIS, however, High-Tech KIS (HTKIS) generates 3.4% of the national synergy. This is more synergy than HTM generates, yet with a much larger number of companies: 193,772 versus 59,621.

Figure 2 shows these specialization patterns by contrasting HTM with HTKIS for the 50+ states. Texas leads in High Tech KIS. Most states do not contribute significantly to either high-tech manufacturing or high-tech KIS. In summary, there are geographical concentrations of high-tech manufacturing and services, but most of the country does not participate significantly at this level of specialization.

*CBSA and CSA*

Of the 940 CBSA distinguished in the data, 446 contribute to the national synergy. Figure 3 shows a map with these CBSA in shades according to their contribution; Figure 4 provides the corresponding map for 139 (of the 165) CSA which contribute to the synergy at this level.[17] The maps show in a bird eye's view that synergy is more concentrated in CSA than CBSA. Not only are the values (expressed as percentages contribution) higher, but the concentration in the north-east (New York-Philadelphia and New England) are more pronounced. The region of LA is clearly indicated in Figure 4, but less so in Figure 3. SF is not indicated as a metropolitan CBSA, but it is as part of the CSA of the Bay area and Silicon Valley.

---

[17] Shapefiles were retrieved from https://www.census.gov/geo/maps-data/data/cbf/cbf_msa.html, adjusted and edited for use in SPSS v22.

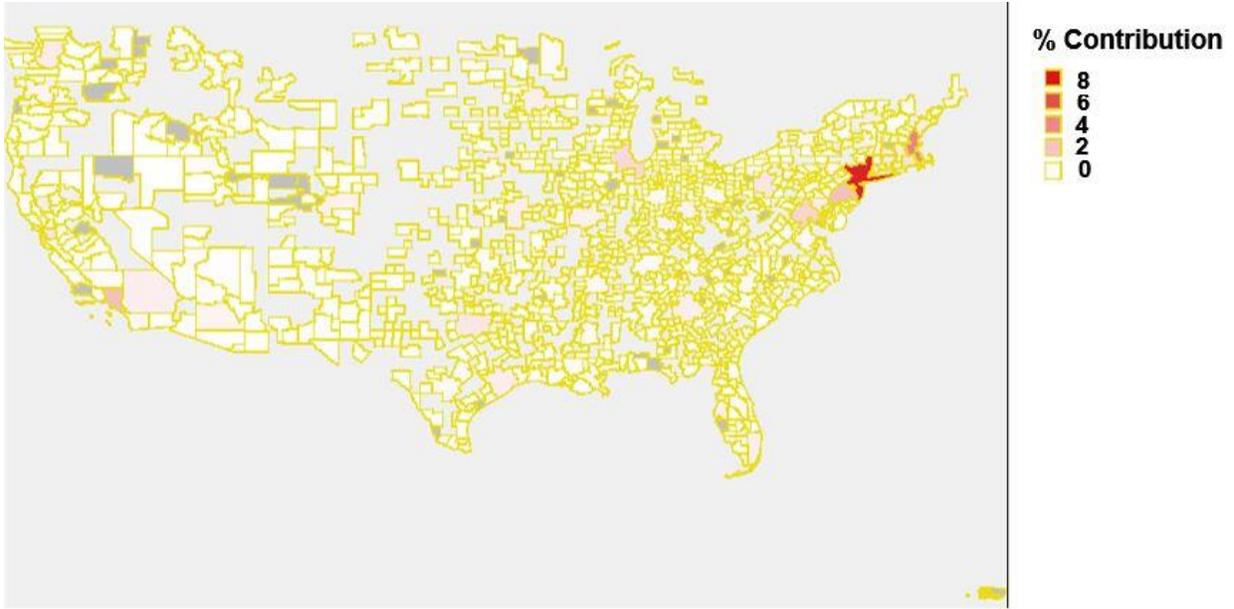

**Figure 3**: Percentages contribution to the national synergy of the USA at the CBSA level.

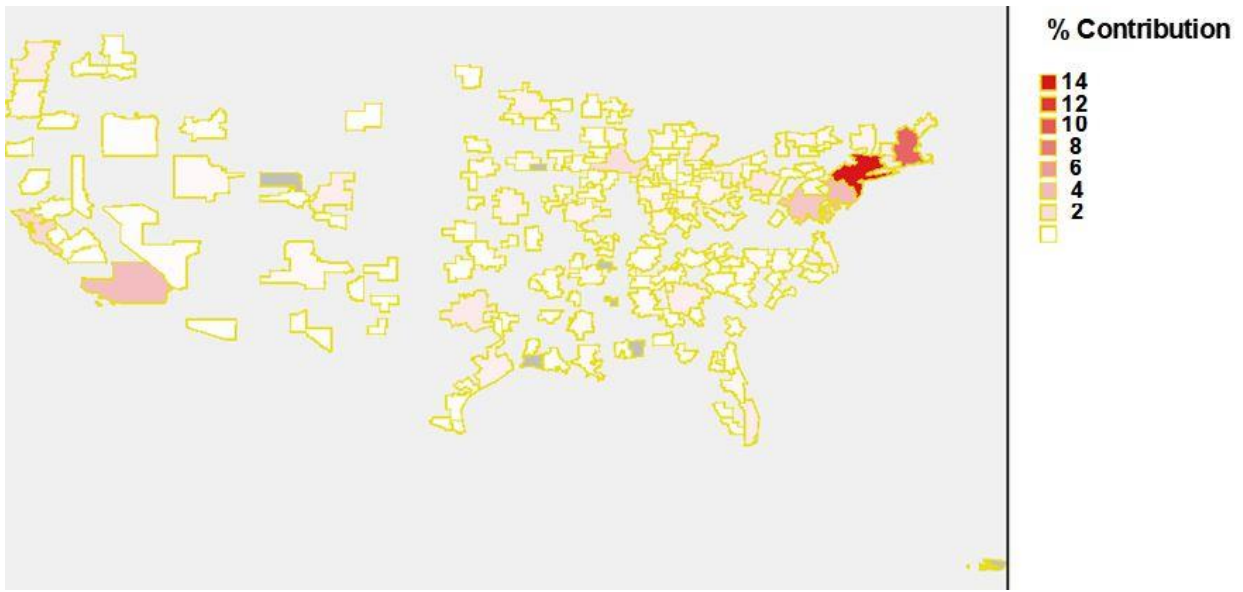

**Figure 4**: Percentages contribution to the national synergy of the USA at the CSA level.

| CBSA | % Contribution | Cumulative % | N of companies | CSA | % Contribution | Cumulative % | N of companies |
|---|---|---|---|---|---|---|---|
| New York-Newark-Jersey City, NY-NJ-PA | 7.52 | 7.52 | 399,754 | New York-Newark, NY-NJ-CT-PA | 13.94 | 13.94 | 478,713 |
| Boston-Cambridge-Newton, MA-NH | 4.17 | 11.69 | 99,429 | Boston-Worcester-Providence, MA-RI-NH-CT | 9.24 | 23.18 | 160,959 |
| Los Angeles-Long Beach-Anaheim, CA | 2.15 | 13.84 | 370,889 | Philadelphia-Reading-Camden, PA-NJ-DE-MD | 3.97 | 27.15 | 138,195 |
| Philadelphia-Camden-Wilmington, PA-NJ-DE-MD | 2.12 | 15.96 | 115,702 | Los Angeles-Long Beach, CA | 3.86 | 31.01 | 463,026 |
| Washington-Arlington-Alexandria, DC-VA-MD-WV | 1.46 | 17.42 | 131,959 | Washington-Baltimore-Arlington, DC-MD-VA-WV-PA | 3.36 | 34.37 | 200,411 |
| Chicago-Naperville-Elgin, IL-IN-WI | 1.28 | 18.70 | 225,971 | San Jose-San Francisco-Oakland, CA | 2.32 | 36.69 | 220,135 |
| Providence-Warwick, RI-MA | 0.93 | 19.63 | 29,223 | Chicago-Naperville, IL-IN-WI | 1.98 | 38.67 | 234,157 |
| San Francisco-Oakland-Hayward, CA | 0.90 | 20.53 | 124,632 | Pittsburgh-New Castle-Weirton, PA-OH-WV | 1.43 | 40.10 | 56,977 |
| Pittsburgh, PA | 0.78 | 21.31 | 50,927 | Hartford-West Hartford, CT | 1.30 | 41.40 | 23,686 |
| Dallas-Fort Worth-Arlington, TX | 0.76 | 22.07 | 208,509 | Dallas-Fort Worth, TX-OK | 1.24 | 42.64 | 217,062 |
| New Haven-Milford, CT | 0.76 | 22.83 | 19,055 | Miami-Fort Lauderdale-Port St. Lucie, FL | 1.11 | 43.75 | 275,500 |
| Hartford-West Hartford-East Hartford, CT | 0.74 | 23.57 | 19,745 | Seattle-Tacoma, WA | 1.07 | 44.82 | 112,394 |
| Worcester, MA-CT | 0.67 | 24.24 | 13,808 | Detroit-Warren-Ann Arbor, MI | 1.05 | 45.87 | 119,770 |
| Baltimore-Columbia-Towson, MD | 0.66 | 24.90 | 53,632 | Cleveland-Akron-Canton, OH | 1.05 | 46.92 | 88,865 |
| Springfield, MA | 0.64 | 25.54 | 13,532 | Denver-Aurora, CO | 1.00 | 47.92 | 89,844 |
| Seattle-Tacoma-Bellevue, WA | 0.63 | 26.17 | 93,594 | Atlanta-Athens-Clarke County-Sandy Springs, GA | 0.97 | 48.89 | 158,547 |
| Bridgeport-Stamford-Norwalk, CT | 0.62 | 26.79 | 24,115 | Portland-Lewiston-South Portland, ME | 0.92 | 49.81 | 13,235 |
| Detroit-Warren-Dearborn, MI | 0.62 | 27.41 | 83,464 | Minneapolis-St. Paul, MN-WI | 0.90 | 50.71 | 81,830 |
| Miami-Fort Lauderdale-West Palm Beach, FL | 0.57 | 27.98 | 257,717 | Houston-The Woodlands, TX | 0.86 | 51.57 | 183,246 |
| Atlanta-Sandy Springs-Roswell, GA | 0.56 | 28.54 | 146,458 | Virginia Beach-Norfolk, VA-NC | 0.85 | 52.42 | 34,552 |
| | | | 2,482,115 | | | | 3,341,104 |

**Table 6**: 20 CBSA (left-column) and CSA (right column) contributing most to the national synergy.

Table 6 lists the top-20 CBSA and CSA in terms of contributions to the synergy in these two domains. The ranking is relatively robust. The sectorial decomposition, however, nuances the picture. Figure 5 shows that the metropolitan regions of New York and Boston deviate by having no synergy contributions from medium-high-tech manufacturing. LA excels in HTM, but is also strongest in MHTM. The (Spearman) rank-order correlations in Table 7 indicate that the ranks vary among sectors.[18] This variation is not among the highest rankings (Table 8), but in the middle range.

**Table 7**: Spearman (rank-order) correlations of percentages of contributions over 900+ CBSA; all correlations are significant at the 1% level.

|         | % all | % htm | % mhtm | % htkis |
|---------|-------|-------|--------|---------|
| % htm   | .415  |       |        |         |
| % mhtm  | .703  | .518  |        |         |
| % htkis | .667  | .541  | .673   |         |
| % kis   | .969  | .421  | .713   | .675    |

Nevertheless, the NY-NJ-PA district makes the most significant contribution to the national synergy (8.65%) among all CBSA (Figure 5). This confirms Feldman & Florida's (1994) observation about the contribution of New Jersey to the national geography of innovation.

---

[18] The Pearson correlations among the four sectorial groups are all above .99.

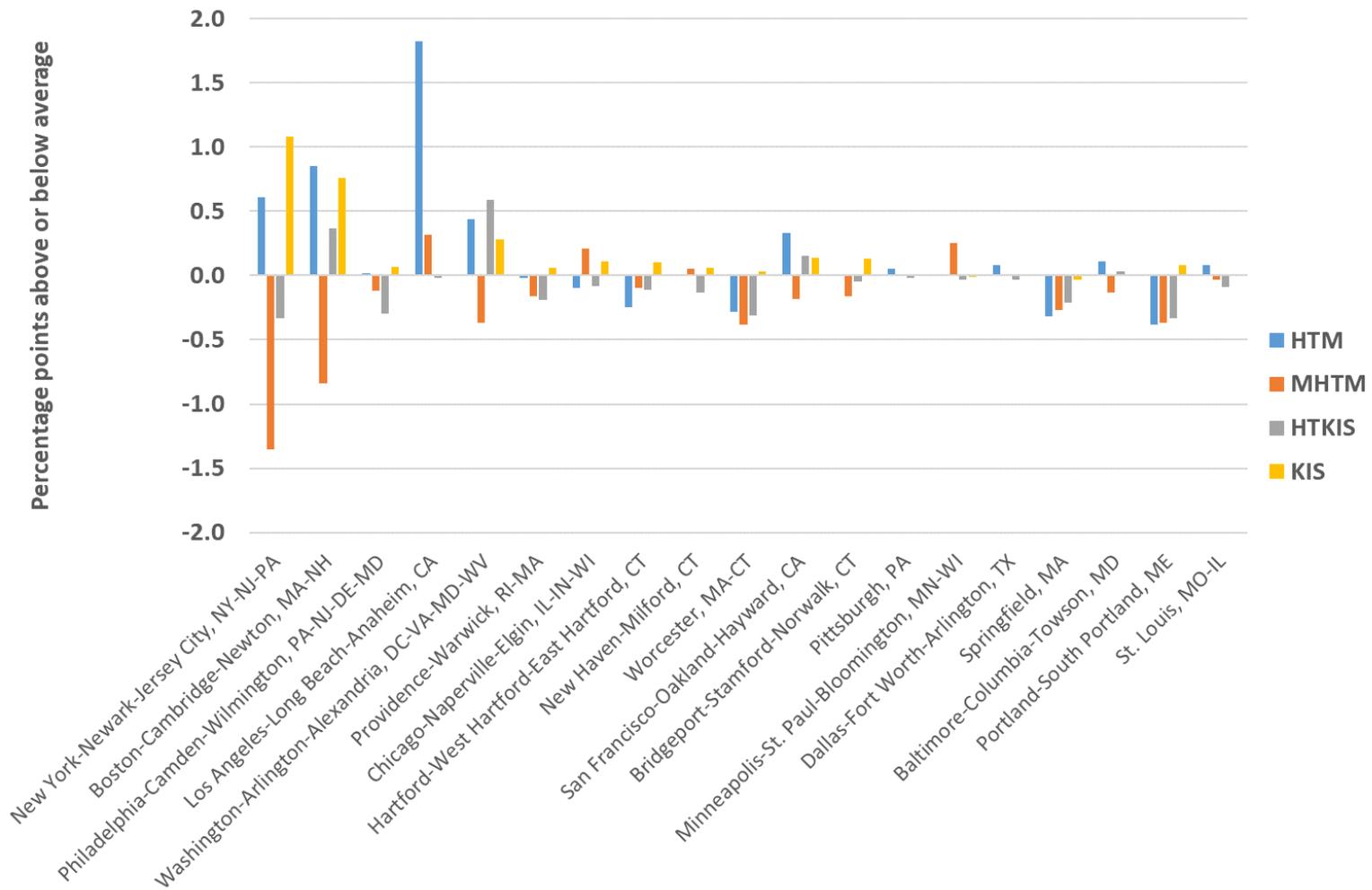

**Figure 5:** Decomposition of the top-20 CBSA in terms of their specialization.

| High-Tech Manufacturing | % | N | High-tech Knowledge-Intensive Services | % | N |
|---|---|---|---|---|---|
| New York-Newark-Jersey City, NY-NJ-PA | 8.02 | 2,861 | New York-Newark-Jersey City, NY-NJ-PA | 7.32 | 12,722 |
| Boston-Cambridge-Newton, MA-NH | 4.81 | 1,238 | Boston-Cambridge-Newton, MA-NH | 4.59 | 3,907 |
| Los Angeles-Long Beach-Anaheim, CA | 3.28 | 3,604 | Los Angeles-Long Beach-Anaheim, CA | 2.24 | 12,532 |
| Washington-Arlington-Alexandria, DC-VA-MD-WV | 1.94 | 1,048 | Washington-Arlington-Alexandria, DC-VA-MD-WV | 2.05 | 7,984 |
| Philadelphia-Camden-Wilmington, PA-NJ-DE-MD | 1.91 | 881 | Philadelphia-Camden-Wilmington, PA-NJ-DE-MD | 1.82 | 2,863 |
| San Francisco-Oakland-Hayward, CA | 1.29 | 1,329 | Chicago-Naperville-Elgin, IL-IN-WI | 1.18 | 5,966 |
| Chicago-Naperville-Elgin, IL-IN-WI | 1.21 | 1,502 | San Francisco-Oakland-Hayward, CA | 1.06 | 5,120 |
| Dallas-Fort Worth-Arlington, TX | 0.85 | 1,568 | Providence-Warwick, RI-MA | 0.83 | 518 |
| Seattle-Tacoma-Bellevue, WA | 0.82 | 908 | Dallas-Fort Worth-Arlington, TX | 0.72 | 6,765 |
| Miami-Fort Lauderdale-West Palm Beach, FL | 0.81 | 1,941 | Pittsburgh, PA | 0.70 | 1,133 |
| Detroit-Warren-Dearborn, MI | 0.79 | 596 | Seattle-Tacoma-Bellevue, WA | 0.68 | 2,826 |
| New Haven-Milford, CT | 0.78 | 170 | Baltimore-Columbia-Towson, MD | 0.68 | 1,635 |
| Pittsburgh, PA | 0.76 | 293 | Virginia Beach-Norfolk-Newport News, VA-NC | 0.64 | 644 |
| Riverside-San Bernardino-Ontario, CA | 0.75 | 619 | Hartford-West Hartford-East Hartford, CT | 0.62 | 416 |
| Providence-Warwick, RI-MA | 0.73 | 171 | New Haven-Milford, CT | 0.62 | 365 |
| Baltimore-Columbia-Towson, MD | 0.72 | 387 | Houston-The Woodlands-Sugar Land, TX | 0.57 | 4,898 |
| Houston-The Woodlands-Sugar Land, TX | 0.63 | 1,151 | Detroit-Warren-Dearborn, MI | 0.57 | 1,978 |
| Bridgeport-Stamford-Norwalk, CT | 0.62 | 170 | Bridgeport-Stamford-Norwalk, CT | 0.53 | 754 |
| Tampa-St. Petersburg-Clearwater, FL | 0.58 | 627 | Miami-Fort Lauderdale-West Palm Beach, FL | 0.53 | 6,677 |

**Table 8**: Top-20 CBSA in High-Tech Manufacturing and High-tech Knowledge-Intensive Services.

The sum of the synergies within the CBSA domain ($N$ = 6,950,681) is 56.8%; 43.2% of the synergy in this domain is realized among the CBSAs. In other words, CBSAs are weakly integrating technologies, markets, and services. They spill-over. One may wish for policy reasons to consider these administratively defined regions as relevant innovation systems, but this claim is not supported by our results. However, the sum of the synergies at the CSA level ($N$ = 5,439,773) is 75.5%; 24.6% of the synergy in this comain is realized above the CSA level. As one would expect, CSAs are integrating technologies, markets, and services to a larger extent than CBSA. The value of 24.6% is of the order of magnitude as the ones found for national systems in Europe. Let us take a closer look to how this works at the level of the state in the case of California.

*California*

Figures 6a and 6b show the specialization patterns of HTM and HTKIS projected on the map of California, respectively. HTM contributes synergy to the LA region, whereas HTKIS provides synergy mainly to the region of San Francisco. The Valley ("San Jose-Sunnyvale-Santa Clara, CA"; $N$ = 49,570) follows at the fifth position with a contribution from both 1,442 HTM companies and 3,014 HTKIS companies. Figure 7 shows the opposition of LA and SF in terms of specializations. The large majority of Californian CBSAs cannot be considered as regional innovation systems generating synergy.

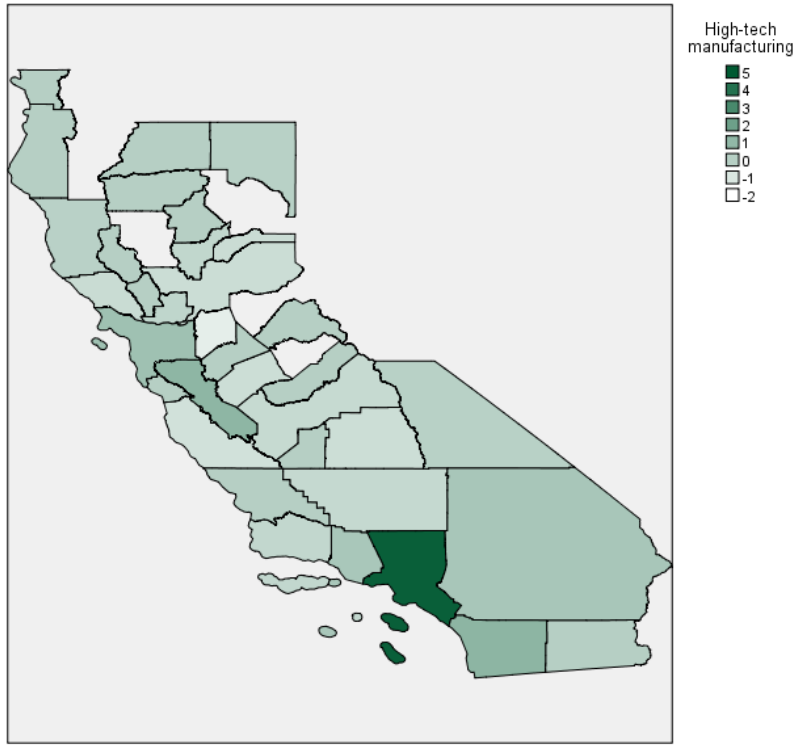 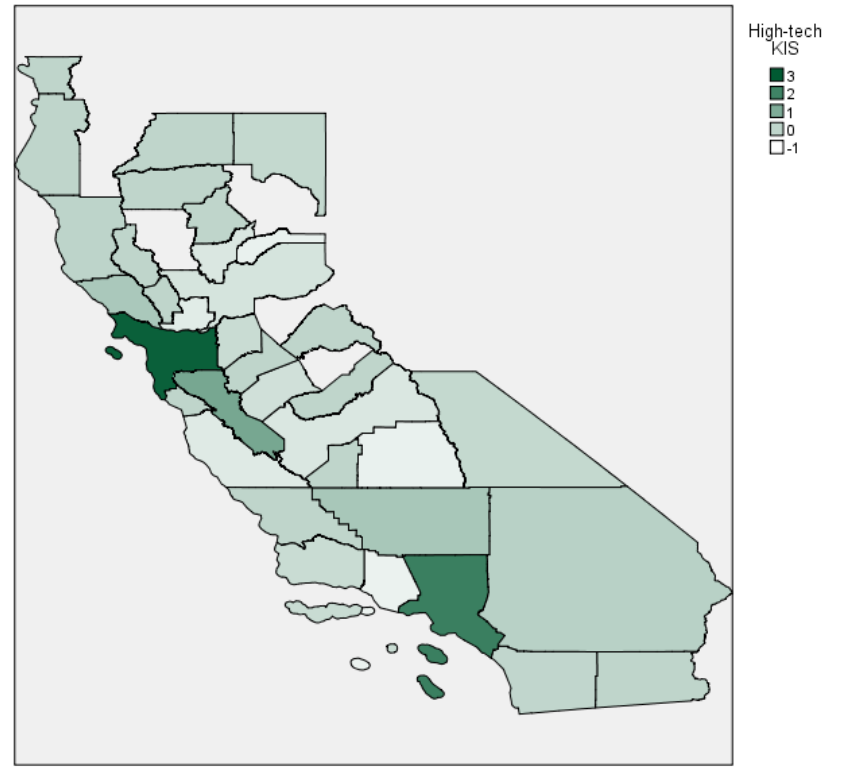

**Figure 6a:** Above and below average contribution of High-Tech Manufacturing to the Synergy in the Knowledge Base of California.

**Figure 6b:** Above and below average contribution of High-Tech KIS to the Synergy in the Knowledge Base of California.

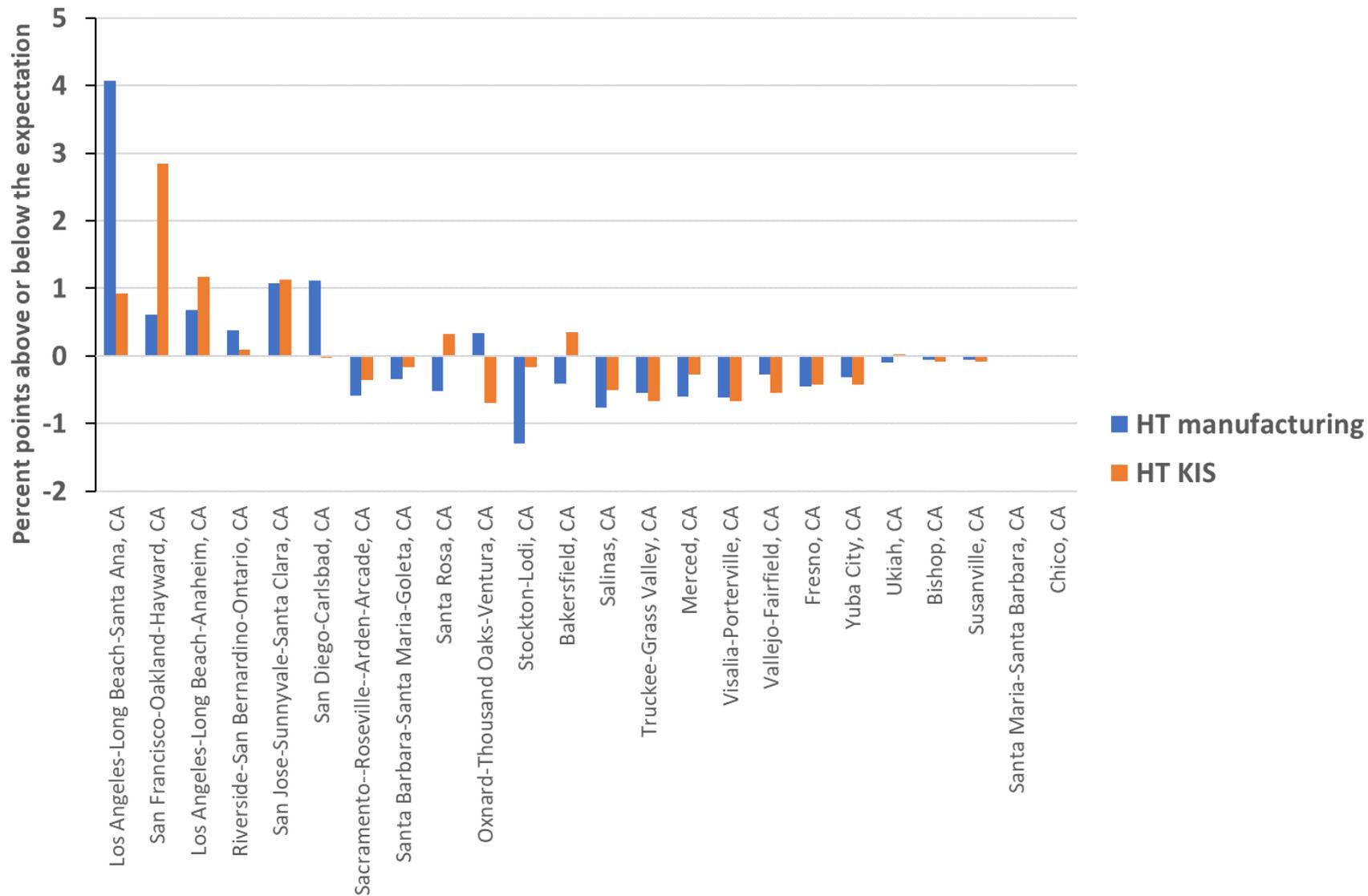

**Figure 7**: Percentages of contributions of high-tech sectors to synergy in Californian regions.

*Silicon Valley and the Bay area*

Silicon Valley is located southeast of the San Francisco area. The CSA of the Bay area is named "San Jose-San Francisco-Oakland, CA" and is composed of seven CBSA. "San Jose-Sunnyvale-Santa Clara, CA" is the CBSA which covers Silicon Valley itself.

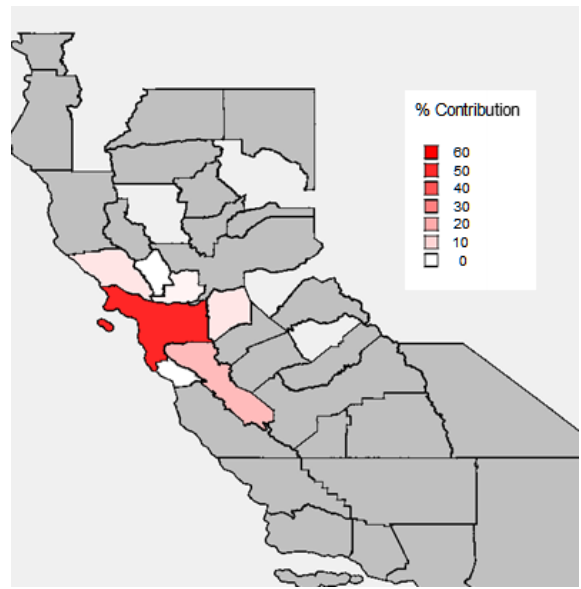

**Figure 8**: Contributions of seven CBSA to the synergy of the CSA "San Jose-San Francisco-Oakland, CA"

**Table 9**: Geographical and sectorial decomposition of the synergy in the CSA "San Jose-San Francisco-Oakland, CA".

| *% contribution to the synergy* | | all sectors | htm | mhtm | htkis | kis |
|---|---|---|---|---|---|---|
| Napa, CA | | 0.00 | 0.00 | 0.00 | 0.00 | 0.00 |
| San Francisco-Oakland-Hayward, CA | | 50.97 | 56.83 | 45.36 | 52.73 | 54.34 |
| San Jose-Sunnyvale-Santa Clara, CA | | 15.65 | 21.10 | 16.96 | 16.94 | 17.16 |
| Santa Cruz-Watsonville, CA | | 0.00 | 0.00 | 0.00 | 0.00 | 0.00 |
| Santa Rosa, CA | | 5.90 | 3.94 | 7.37 | 6.45 | 5.47 |
| Stockton-Lodi, CA | | 5.73 | 0.43 | 6.13 | 4.22 | 4.67 |
| Vallejo-Fairfield, CA | | 2.38 | 1.31 | 0.96 | 0.10 | 1.67 |
| | Sum | 80.63 | 83.61 | 76.78 | 80.44 | 83.31 |
| | $T_0$ | 19.37 | 16.39 | 23.22 | 19.56 | 16.69 |
| *n of companies* | | all sectors | htm | mhtm | htkis | kis |
| Napa, CA | | 3,905 | 17 | 64 | 59 | 1,230 |
| San Francisco-Oakland-Hayward, CA | | 124,632 | 1,329 | 1,799 | 5,120 | 51,285 |
| San Jose-Sunnyvale-Santa Clara, CA | | 49,493 | 1,439 | 1,159 | 3,010 | 20,094 |
| Santa Cruz-Watsonville, CA | | 7,707 | 80 | 141 | 224 | 2,809 |
| Santa Rosa, CA | | 14,613 | 121 | 279 | 301 | 4,978 |
| Stockton-Lodi, CA | | 12,538 | 48 | 242 | 144 | 3,638 |
| Vallejo-Fairfield, CA | | 7,247 | 46 | 125 | 136 | 2,300 |
| | Sum | 220,135 | 3,080 | 3,809 | 8,994 | 86,334 |
| per sector | | | | | | |
| % N of companies | | 46.43 | 1.40 | 1.73 | 4.09 | 39.22 |
| % contribution | | 44.82 | 3.43 | 6.16 | 3.97 | 31.27 |

More than 80% of the synergy in the CSA is generated by the seven CBSA; that is, within the area. The contributions to the synergy are not sector-specific. However, the last lines of Table 9 teach us that 1.40% of the companies in HTM contribute 3.43% to the synergy in this region, whereas HTKIS and KIS contribute proportionally less than expected.

Let us further decompose. The data of the CBSA "San Jose-Sunnyvale-Santa Clara, CA" contains two zip-codes at the two-digit level: 94 and 95. Companies with zipcode 95 (n = 36,330) generate 43.27% of the synergy in this CBSA; 3.95% is generated by 13,184 companies with zipcode 94. Further decomposition of the Valley is possible in terms of cities or companies. Using company names, however, one obtains maximum entropy in the geographical dimension because all companies have unique names. Decomposition in terms of cities leads to subsets which have the city as a constant and consequently zero entropy in the geographical dimension. In the latter case, the redundancy is necessarily zero and in the former model uncertainty prevails to such an extent that $T_{UIG}$ is part of the maximum entropy and therefore necessarily positive ($T_{UIG} > 0$). In both these cases, no synergy can be measured for methodological reasons. In other words, this methodology cannot be used for the lowest level because there is either no variance in a single city name or maximum entropy when using unique company names.

In Silicon Valley (SV), HTM contributes more to the synergy than MHTM and HTKIS more than KIS, although both KIS and HTKIS contribute less than expected. HTKIS, for example, contributes 4.05% to the synergy with 6.06% of the companies, whereas HTM contributes 5.20% with 2.90% of the companies. In other words, there is much more KIS (and HTKIS) than HTM and MHTM in Silicon Valley, but the synergy contribution of manufacturing is much higher than that of the knowledge-intensive services.

**Discussion and limitations**

We used synergy among the distributions of sectors, company addresses, and size-classes as an indicator of innovation-systemness to study the U.S. at various levels of aggregation. Obviously, the main limitation of this study is the use of ORBIS data. We have no access to how the data is collected; the database is private property. Still, it is probably the best data currently available for this type of study. As noted, ZIP codes vary over geographical regions; however, in reference to the other two dimensions, the distribution of ZIP codes indicates local constraints (such as infrastructure) operating as a selection environment. In the case of NACE codes, the alternative of NIAC would be an option, and other schemes could be used for the scaling of companies in terms of size-classes. Most importantly, the definition of what counts as a company in the database is beyond our control.

The results thus provide us only with a window, and we do not wish to deny that other approaches are possible and perhaps even more fruitful. However, we improve on other approaches by moving beyond a political definition of innovation systems to an empirical one which can be tested by using synergy as a measure of systemness (cf. Griliches, 1994). Systemness can then also be rejected as a fruitful hypothesis at specific levels of aggregation and/or within sectors. Our results, for example, do not indicate the systemness of the U.S. innovation system at the national level.

Given the proviso of the methodological constraints of the study, our analysis suggests that the states, and not the nation or regions, are the most relevant innovation environments in the U.S. To the extent that states are the relevant geographical entities for innovation, significant policy considerations should follow. Note that our conclusions do not imply volition or initiative on the part of state governments; these are input data. We are just reporting empirical findings about systemic configurations. In the past, state initiatives have often been evaluated as ineffective or incompetent compared with initiatives at the level of metropolitan regions (Agrawal, Cockburn, Galasso, & Oettl, 2014; Bartik, 2017). Even so, states have a long history of creating baskets of incentives, training, and investment programs to grow industry (Shapira & Youtie, 2010). Our results indicate synergy in the knowledge base of specific states along the East Coast (New Jersey, Massachusetts, New York, and Pennsylvania), in California, and, in the case of HTKIS, Texas.

The regions measured as CBSAs are too small to comprise innovation systems; the innovation systems spill over the boundaries of these units of analysis. As could be expected, CSAs—combining contiguous CBSAs—are more appropriate units of analysis in terms of the development of synergy. The decomposition in terms of sectors shows specialization among states and regions, but does not change the main pattern other than modulating it. The overall picture is one of concentration of high-tech and dispersed specialization at many different locations. Knowledge-intensive services are dominant, but do not contribute to the synergy above expectation.

Focusing on California, three regions are most relevant for the discussion: LA with synergy in manufacturing (both HTM and MHTM), San Francisco with synergy in KIS and HTKIS, and Silicon Valley with mainly KIS in the portfolio but manufacturing as the generator of synergy. The services in Silicon Valley are not contributing to synergy in the region but operating at national and global levels. While these conclusions may not be surprising from the perspective of hindsight, *ex ante* it would have been difficult to specify the nuances in such detail without a quantitative analysis.

**Acknowledgement**

The authors are grateful to Inga Ivanova, Dieter Franz Kogler, and the anonymous referees for their comments. IP acknowledges financial support from the Basque Government's Department of Education, Language Policy and Culture (Grant Number IT885-16).

**References**

Agrawal, A., Cockburn, I., Galasso, A., & Oettl, A. (2014). Why are some regions more innovative than others? The role of small companies in the presence of large labs. *Journal of Urban Economics, 81*, 149-165.

Audretsch, D. B., & Feldman, M. P. (1996). R&D spillovers and the geography of innovation and production. *The American Economic Review, 86*(3), 630-640.

Bartik, T. J. (2017). *A New Panel Database on Business Incentives for Economic Development Offered by State and Local Governments in the United States.* Kalamazoo, MI: W.E. Upjohn Institute for Employment Research. Retrieved from http://research.upjohn.org/cgi/viewcontent.cgi?article=1228&context=reports on October 1, 2017

Blau, P. M., & Schoenherr, R. A. (1971). *The Structure of Organisations*. New York: Basic Books.

Boschma, R., Balland, P.-A., & Kogler, D. F. (2014). Relatedness and technological change in cities: the rise and fall of technological knowledge in US metropolitan areas from 1981 to 2010. *Industrial and Corporate Change, 24*(1), 223-250.

Bresnahan, T.F. & Gambardella, A. (eds.) (2004). *Building High-Tech Clusters. Silicon Valley and Beyond.* New York: Cambridge University Press.

Brooks, D. R., & Wiley, E. O. (1986). *Evolution as Entropy*. Chicago/London: University of Chicago Press.

Brown, D. L., Cromartie, J. B., & Kulcsar, L. J. (2004). Micropolitan areas and the measurement of American urbanization. *Population Research and Policy Review*, *23*(4), 399-418.

Bruckner, E., Ebeling, W., Montaño, M. J., & Scharnhorst, A. (1996). Nonlinear stochastic effects of substitution—an evolutionary approach. *Journal of Evolutionary Economics, 6*(1), 1-30.

Carlsson, B. (2006). Internationalization of Innovation Systems: A Survey of the Literature. *Research Policy, 35*(1), 56-67.

Carlsson, B. (2013). Knowledge Flows in High-Tech Industry Clusters: Dissemination Mechanisms and Innovation Regimes. In E. S. Andersen & A. Pyka (Eds.), *Long Term Economic Development: Demand, finance, organization, policy and innovation in Schumpeterian perspective* (pp. 191-222). Berlin / Heidelberg: Springer Verlag.

Carter, A. P. (1996). Measuring the performance of a knowledge-based economy. In D. Foray & B.-Å. Lundvall (Eds.), *Employment and growth in the knowledge-based economy* (pp. 61-68). Paris: OECD.

Casson, M. (1997). *Information and Organization: A New Perspective on the Theory of the Firm*. Oxford: Clarendon Press.

Cooke, P. (2002). *Knowledge Economies*. London: Routledge.

Cucco, I., & Leydesdorff, L. (2013; in preparation). Measuring synergies in the Italian innovation system: A comparison of Amadeus™ and administrative data.

Eurostat (2008). *NACE Rev. 2: Statistical classification of economic activites in the European Community*. Luxembourg: Office for Official Publications of the European Communities; retrieved from https://www.economy.com/support/blog/getfile.asp?did=5E5D88CF-81DEF-4655-8767-ADF4601EC4463D4650&fid=cc4033c4650c4672be4654a4653e8035a4658b4459ee4650de4650.pdf, 30 September 2018.

Feldman, M. P., & Florida, R. (1994). Innovation in the United States. *Annals of the Association of American Geographers*, *84*(2), 210-229.

Feldman, M., & Storper, M. (2016). Economic growth and economic development: geographic dimensions, definitions and disparities. *The New Handbook of Economic Geography. Oxford University Press: Oxford*.

Florida, R. (2002). Bohemia and economic geography. *Journal of Economic Geography, 2*(1), 55-71.

Freeman, C. (1987). *Technology and Economic Performance: Lessons from Japan*. London: Pinter.

Freeman, C., & Soete, L. (1997). *The Economics of Industrial Innovation* (Third Edition). London: Pinter.

Geels, F. W., & Schot, J. (2007). Typology of sociotechnical transition pathways. *Research Policy, 36*(3), 399-417.

Godin, B. (2006). The Knowledge-Based Economy: Conceptual Framework or Buzzword? *Journal of Technology Transfer, 31*(1), 17-30.

Griliches, Z. (1994). Productivity, R&D and the Data constraint. *American Economic Review, 84*(1), 1-23.

Hall, P. (2009). Looking backward, looking forward: the city region of the mid-21st century. *Regional Studies*, *43*(6), 803-817.

Hodgson, G., & Knudsen, T. (2011). *Darwin's Conjecture: The Search for General Principles of Social and Economic Evolution*. Chicago / London: University of Chicago Press.

Ivanova, I. A., & Leydesdorff, L. (2014). Rotational Symmetry and the Transformation of Innovation Systems in a Triple Helix of University-Industry-Government Relations. *Technological Forecasting and Social Change, 86*, 143-156.

Jacobs, J. (1961). *The Death and Life of Great American Cities*. New York, NY: Vintage Books.

Krippendorff, K. (2009). Information of Interactions in Complex Systems. *International Journal of General Systems, 38*(6), 669-680.

Langton, C. G. (1989). Artificial Life. In C. G. Langton (Ed.), *Artificial Life* (Vol. VI, pp. 1-47). Redwood, etc.: Addison-Wesley.

Lengyel, B., & Leydesdorff, L. (2011). Regional innovation systems in Hungary: The failing synergy at the national level. *Regional Studies, 45*(5), 677-693. doi: DOI: 10.1080/00343401003614274

Leydesdorff, L. (2000). Is the European Union Becoming a Single Publication System? *Scientometrics, 47*(2), 265-280.

Leydesdorff, L., & Ahrweiler, P. (2014). In search of a network theory of innovations: Relations, positions, and perspectives. *Journal of the Association for Information Science and Technology, 65*(11), 2359-2374.

Leydesdorff, L., & Fritsch, M. (2006). Measuring the Knowledge Base of Regional Innovation Systems in Germany in Terms of a Triple Helix Dynamics. *Research Policy, 35*(10), 1538-1553.

Leydesdorff, L., & Porto-Gomez, I. (early view 2017). Measuring the Expected Synergy in Spanish Regional and National Systems of Innovation. *Journal of Technology Transfer*. doi: 10.1007/s10961-017-9618-4

Leydesdorff, L., & Strand, Ø. (2013). The Swedish System of Innovation: Regional Synergies in a Knowledge-Based Economy. *Journal of the American Society for Information Science and Technology, 64*(9), 1890-1902; doi: 1810.1002/asi.22895.

Leydesdorff, L., & Zhou, P. (2014). Measuring the Knowledge-Based Economy of China in terms of Synergy among Technological, Organizational, and Geographic Attributes of Companies. *Scientometrics, 98*(3), 1703-1719. doi: 10.1007/s11192-013-1179-1

Leydesdorff, L., Dolfsma, W., & Van der Panne, G. (2006). Measuring the Knowledge Base of an Economy in Terms of Triple-Helix Relations among 'Technology, Organization, and Territory'. *Research Policy, 35*(2), 181-199.

Leydesdorff, L., Park, H. W., & Lengyel, B. (2014). A routine for measuring synergy in university–industry–government relations: Mutual information as a Triple-Helix and Quadruple-Helix indicator. *Scientometrics, 99*(1), 27-35.

Leydesdorff, L., Perevodchikov, E., & Uvarov, A. (2015). Measuring triple-helix synergy in the Russian innovation systems at regional, provincial, and national levels. *Journal of the Association for Information Science and Technology, 66*(6), 1229-1238.

Lundvall, B.-Å. (Ed.). (1992). *National Systems of Innovation*. London: Pinter.

McGill, W. J. (1954). Multivariate information transmission. *Psychometrika, 19*(2), 97-116.

Mowery, D. C., & Rosenberg, N. (1979). The influence of market demand upon innovation: a critical review of some empirical studies. *Research Policy, 8*, 102-153.

Nelson, R. R. (Ed.). (1993). *National Innovation Systems: A Comparative Analysis*. New York: Oxford University Press.

Nelson, R. R., & Winter, S. G. (1977). In Search of Useful Theory of Innovation. *Research Policy, 6*(1), 35-76.

Petersen, A., Rotolo, D., & Leydesdorff, L. (2016). A Triple Helix Model of Medical Innovations: *Supply*, *Demand*, and *Technological Capabilities* in Terms of Medical Subject Headings. *Research Policy, 45*(3), 666-681. doi: 10.1016/j.respol.2015.12.004

Pugh, D. S., Hickson, D. J., & Hinings, C. R. (1969). An empirical taxonomy of structures of work organizations. *Administrative Science Quarterly*, 115-126.

Pugh, D. S., Hickson, D. J., Hinings, C. R., & Turner, C. (1969). The context of organization structures. *Administrative Science Quarterly*, 91-114.

Rocha, F. (1999). Inter-company technological cooperation: effects of absorptive capacity, firm-size and specialization. *Economics of Innovation and New Technology*, *8*(3), 253-271.

Saxenian, A. (1996). *Regional Advantage: Culture and Competition in Silicon Valley and Route 128*. Cambridge, MA: Harvard University Press.

Shannon, C. E. (1948). A Mathematical Theory of Communication. *Bell System Technical Journal, 27 (July and October)*, 379-423 and 623-656.

Shapira, P., & Youtie, J. (2010). The Innovation System and Innovation Policy in the United States. In R. Frietsch & M. Schüller (Eds.), *Competing for Global Innovation Leadership: Innovation Systems and Policies in the USA, EU and Asia* (Chapter 2, pp. 5-29). Stuttgart: Fraunhofer IRB Verlag.

Storper, M. (1997). *The Regional World – Territorial Development in a Global Economy*. New York: Guilford Press.

Storper, M., Kemeny, T., Makarem, N., Makarem, N. P., & Osman, T. (2015). *The Rise and Fall of Urban Economies*. Stanford, CA: Stanford University Press.

Strand, Ø., & Leydesdorff, L. (2013). Where is Synergy in the Norwegian Innovation System Indicated? Triple Helix Relations among Technology, Organization, and Geography. *Technological Forecasting and Social Change, 80*(3), 471-484.

Theil, H. (1972). *Statistical Decomposition Analysis*. Amsterdam/ London: North-Holland.

Ulanowicz, R. E. (2009). The dual nature of ecosystem dynamics. *Ecological Modelling, 220*(16), 1886-1892.

Whittington, K. B., Owen-Smith, J., & Powell, W. W. (2009). Networks, propinquity, and innovation in knowledge-intensive industries. *Administrative Science Quarterly*, *54*(1), 90-122.

**Appendix 1:** Descriptive statistics of the numbers of companies in U.S. states.

|  | All sectors | High-Tech Manufacturing (HTM) | Medium-high-tech anufacturing (MHTM) | Knowledge-intensive services (KIS) | High-tech KIS (HTKIS) |
|---|---|---|---|---|---|
| USA | 8,121,301 | 59,621 | 140,594 | 2,789,295 | 193,772 |
| Alaska | 18,859 | 120 | 255 | 6,251 | 444 |
| Alabama | 104,052 | 820 | 1,821 | 31,621 | 1,713 |
| Arkansas | 65,395 | 360 | 1,111 | 19,125 | 921 |
| Arizona | 140,726 | 1,291 | 2,428 | 49,055 | 3,178 |
| California | 951,223 | 10,100 | 16,963 | 350,122 | 29,591 |
| Colorado | 151,701 | 1,276 | 2,458 | 54,565 | 4,414 |
| Connecticut | 110,330 | 862 | 2,104 | 38,953 | 2,475 |
| District of Columbia | 21,752 | 95 | 85 | 11,032 | 1,198 |
| Delaware | 23,071 | 146 | 332 | 7,648 | 487 |
| Florida | 726,524 | 5,061 | 10,292 | 259,870 | 16,338 |
| Georgia | 242,053 | 1,601 | 4,032 | 80,289 | 5,687 |
| Guam | 371 | 1 | 7 | 135 | 8 |
| Hawaii | 26,383 | 118 | 236 | 9,065 | 574 |
| Iowa | 88,436 | 360 | 1,505 | 23,331 | 1,306 |
| Idaho | 44,591 | 454 | 855 | 12,819 | 736 |
| Illinois | 306,798 | 1,837 | 6,006 | 106,217 | 7,086 |
| Indiana | 155,277 | 791 | 3,343 | 46,833 | 2,661 |
| Kansas | 73,392 | 454 | 1,370 | 23,538 | 1,364 |
| Kentucky | 97,134 | 411 | 1,663 | 30,261 | 1,565 |
| Louisiana | 105,732 | 505 | 1,834 | 35,883 | 1,829 |
| Massachusetts | 200,387 | 2,150 | 3,431 | 72,722 | 6,469 |
| Maryland | 145,392 | 1,130 | 1,695 | 54,455 | 5,273 |
| Maine | 34,816 | 219 | 508 | 10,575 | 573 |
| Michigan | 252,780 | 1,556 | 5,728 | 79,214 | 4,764 |
| Minnesota | 161,422 | 1,024 | 3,165 | 48,836 | 3,238 |
| Missouri | 144,676 | 705 | 2,734 | 45,741 | 2,550 |
| Mississippi | 61,900 | 407 | 921 | 18,601 | 872 |
| Montana | 35,033 | 670 | 620 | 10,084 | 605 |
| North Carolina | 218,808 | 1,341 | 3,966 | 66,766 | 4,211 |
| North Dakota | 22,820 | 103 | 385 | 6,089 | 348 |
| Nebraska | 54,723 | 215 | 841 | 15,838 | 856 |
| New Hampshire | 39,308 | 484 | 943 | 12,126 | 979 |
| New Jersey | 235,364 | 1,898 | 3,779 | 83,743 | 6,406 |
| New Mexico | 44,649 | 385 | 678 | 14,795 | 1,019 |
| Nevada | 53,365 | 419 | 925 | 19,158 | 1,382 |
| New York | 520,850 | 3,369 | 6,586 | 185,429 | 13,950 |
| Ohio | 278,319 | 1,709 | 6,585 | 92,000 | 5,296 |
| Oklahoma | 80,655 | 467 | 1,809 | 27,658 | 1,765 |
| Oregon | 121,216 | 1,711 | 2,222 | 37,096 | 2,428 |
| Pennsylvania | 295,967 | 1,927 | 5,564 | 100,155 | 6,063 |
| Puerto Rico | 3,911 | 55 | 90 | 1,418 | 122 |
| Rhode Island | 28,190 | 173 | 467 | 9,216 | 520 |
| South Carolina | 102,707 | 590 | 1,967 | 30,219 | 1,533 |
| South Dakota | 26,224 | 119 | 466 | 6,775 | 366 |
| Tennessee | 140,548 | 730 | 2,568 | 44,232 | 2,343 |
| Texas | 708,057 | 4,585 | 11,951 | 286,159 | 19,245 |
| Utah | 62,131 | 578 | 1,253 | 20,672 | 1,559 |
| Virginia | 176,699 | 1,191 | 2,261 | 63,246 | 6,962 |
| Virgin Islands | 258 | 1 | 3 | 100 | 11 |
| Vermont | 20,624 | 158 | 294 | 6,614 | 444 |
| Washington | 177,391 | 1,703 | 3,026 | 58,389 | 4,249 |
| Wisconsin | 166,981 | 944 | 3,649 | 47,458 | 2,832 |
| West Virginia | 31,721 | 128 | 463 | 11,119 | 553 |
| Wyoming | 19,609 | 114 | 351 | 5,984 | 411 |